\documentclass[12pt]{article}
\usepackage{amssymb}
\usepackage{psfig}
\begin{document}
\begin{center}
\textbf{COSMIC EVOLUTION AND PRIMORDIAL BLACK HOLE EVAPORATION}\\

\bigskip
\bigskip
\bigskip

I. Brevik\footnote{E-mail address: iver.h.brevik@mtf.ntnu.no} and G. Halnes\footnote{E-mail address: geir.halnes@bi.slu.se}\\

\bigskip
\bigskip

Division of Applied Mechanics,\\ Norwegian University of Science and Technology,\\
N-7491 Trondheim, Norway\\

\bigskip
\bigskip


\bigskip
\bigskip

Revised version, November 2002
\end{center}

\bigskip

\begin{abstract}

A cosmological model in which primordial black holes (PBHs) are present in the cosmic fluid at some instant $t=t_0$ is investigated. The time $t_0$ is naturally identified with the end of the inflationary period. The PBHs are assumed to be nonrelativistic in the comoving fluid, to have the same mass, and may be subject to evaporation for $t>t_0$. Our present work is related to an earlier paper of Zimdahl and Pav\'{o}n [Phys. Rev. D {\bf 58}, 103506 (1998)], but in contradistinction to these authors we assume that the (negative) production rate of the PBHs is zero. This assumption appears to us to be more simple and more physical. Consequences of the formalism are worked out. In particular, the four-divergence of the entropy four-vector in combination with the second law in thermodynamics show in a clear way how the the case of PBH evaporation corresponds to a production of entropy. Accretion of radiation onto the black holes is neglected. We consider both a model where two different sub-fluids interact, and a model involving one single fluid only. In the latter case an effective  bulk viscosity naturally appears in the formalism.

\end{abstract}

\newpage

\section{Introduction}

In studies of the early universe, the idea has sometimes been put forward that under these extreme physical conditions primordial black holes (PBHs) were present in the cosmic fluid during some brief time period. When considered in the comoving coordinate system, one may naturally  take these PBHs to be nonrelativistic particles immersed in a cosmic fluid, most likely a radiation dominated fluid.  The situation resembles that of a two-phase system of air bubbles in water, and similarly as in ordinary hydromechanics one may consider this physical system in one of the following two ways: either, as a system of two different fluid components in interaction, or, as an effective one-component fluid endowed with a bulk viscosity (under the assumption of complete isotropy the shear viscosity does not come into play). 

The 1998 paper of Zimdahl and Pav\'{o}n \cite{zimdahl98} (hereafter referred to as ZP; cf. also the more recent Ref.~\cite{zimdahl01}) discussed a two-fluid model of the early universe, namely a nonrelativistic PBH component with particle masses lying within a narrow range, obeying the equation of state for dust, embedded in a radiation dominated cosmic fluid. The influence from the PBHs on the dynamics of the universe was described by the two-fluid model developed earlier by Zimdahl in Ref.~\cite{zimdahl97}. 

A central ingredient of this theory was that the PBHs were allowed to be created, or annihilated, as long as the law of total energy conservation was not broken. One reason why this kind of model is physically appealing, is that the dynamics allows {\it entropy} to be produced. As is well known, one of the challenges of cosmological theory is to account for the large amount of specific entropy observed in the universe; the nondimensional entropy per baryon being about $\sigma \simeq 10^9$ \cite{weinberg71}. (It turns out that, if instead of a two-fluid model one chooses to work in terms of an effective one-component model, the interaction between the PBHs and the radiation fluid can be described by means of an effective bulk viscosity $\zeta$; cf. Eq.~(60) below.)

The purpose of the present paper is to focus attention on the following point. ZP assumed that all the PBHs, with practically equal mass $m_{BH}$, were produced during a very short time interval prior to a a definite instant $t=t_0$. The magnitude of $t_0$ was actually not given explicitly in Ref.~\cite{zimdahl98}, but it seems reasonable to assume that $t_0 \sim 10^{-33}$ s, {\it i.e.,} inflationary times. (In an earlier work of Hayward and Pav\'{o}n \cite{hayward89} the time $t_0$ at which the PBHs were created was pushed further back in the history of the universe, to the order of a few Planck times.) For $t >t_0$, ZP assumed the evolution of the PBH fluid to be governed by a fixed and negative production rate $\Gamma_{ BH}$:
\begin{equation}
\Gamma_{BH}=-\frac{1}{\tau},
\end{equation}
\label{1}
$\tau$ being the assumed constant life time of an evaporating PBH particle which at $t_0$ has a given mass $m_{BH}(t_0)$. The history of the universe after $t=t_0$ is thus in this picture determined by the Friedmann equations, together with the continuity of the PBH particles as determined by Eq.~(\ref{1}).  The decay of the PBHs according to this picture, is analogous 
 to the decay of an assembly of radioactive particles. Thus if one considers a comoving volume element, there is a definite probability for a certain fraction of the PBHs simply to disappear.

It ought to be mentioned that the physical interpretation of $\Gamma_{BH}$ is somewhat problematic. It seems to be most natural to let this quantity refer to {\it entire} BH particles, so that $\Gamma_{BH}$ means  the specific fraction of BH particles disappearing, per unit time and unit comoving volume, from the cosmic fluid. "Specific" here means per particle. In general, if $\Psi$ denotes the production rate of particles per unit comoving volume, one has $(nU^\mu)_{;\mu}=\Psi$, where $U^\mu$ is the four-vector of the cosmic fluid and $n$ the particle density (cf., for instance, Ref.~\cite{brevik96}). As $\Psi=n\Gamma$, one thus has the general equation
\begin{equation}
(nU^\mu)_{;\mu}=n\Gamma,
\end{equation}
\label{2} 
showing explicitly how the quantity $\Gamma$ refers to one particle. It may be instructive to compare with the interpretation of the Hoyle-Narlikar scalar C - field (cf, for instance, Ref.~\cite{hawking73}): in that case, a continuous creation of matter is postulated to take place everywhere at any time; in the present case, a continuous destruction of BH matter may similarly be postulated to occur, during the life time $\tau$.

One may however wonder: is this way of looking at the PBH evaporation physically correct? Imagine how the evaporation of the PBHs takes place: The black hole mass $m_{BH}(t)$ changes with time, both because of absorption of electromagnetic energy (accretion), and because of Hawking radiation, but there appears to be no natural physical mechanism whereby some of the black hole particles should disappear suddenly. It may be argued that it would be more natural to assume that one black hole will not be able to disappear if not all the others do. And this is our main opinion to be forwarded in this paper: Assuming homogeneity and equal mass of all the black holes, we argue that it is physically more plausible to assume that all the PBHs decay in exactly the same way, having the same life time $\tau$. Mathematically, this amounts to setting the negative production rate of the PBHs equal to zero:
\begin{equation}\Gamma_{BH}=0.
\end{equation}
\label{3}
The theme of our treatment in the following is thus to show how the universe develops after the instant $t=t_0$ if the condition (3) is adopted, instead of the ZP condition (\ref{1}). Therewith we avoid the problematic interpretation of $\Gamma_{BH}$ altogether. No production or destruction of black holes is permitted. The formalism given by ZP will have to be changed and simplified. We will work out the changed  formalism both for a two-fluid model, and for a one-fluid model. As already anticipated, the one-fluid model turns out to lead to an entropy production via an effective bulk viscosity, in a natural way.

Before embarking upon an analysis of the two-fluid model, let us briefly consider the decay law for a single evaporating black hole. We assume that at time $t_0$  all the PBHs with the same mass $m_{BH}(t_0)$ have been produced, and start evaporating. From Stefan-Boltzmann's law we have $dm_{BH}/dt=-\tilde{\sigma}T_{BH}^4\,A$, where $\tilde{\sigma}$ is Stefan-Boltzmann's constant  and $A=16\pi G^2 m_{BH}^2$ is the surface area of a black hole with Schwarzschild radius $2 G m_{BH}$. Integration with respect to mass from the instant $t_0$ onwards yields
\begin{equation}
m_{BH}(t)=\left( m_{BH}^3(t_0)-3C(t-t_0)\right)^{1/3},
\end{equation}
\label{4}
where $C$ is a constant.
Since we associate all the PBHs with the same life time $\tau$, we require that $m_{BH}=0$ at $t=t_0+\tau$. That determines the decay law; for later convenience we write it in the form
\begin{equation}
\frac{\dot{m}_{BH}(t)}{m_{BH}(t)}=\frac{-1}{3\tau}\left( 1-\frac{t-t_0}{\tau}\right)^{-1}.
\end{equation}
\label{5}
The following point should be emphasized: we will ignore the effect of {\it accretion} onto the black holes. This assumption puts a constraint on the magnitude of $\tau$, as can be seen from a simple argument: When a Schwarzschild black hole is immersed in an isotropic bath of photons with energy density $\rho_{rad}$ far from the black hole, the accretion rate $\dot{m}_{BH}(t)$ is known to be \footnote{ This argument ignores the universe expansion.}
\begin{equation}
\dot{m}_{BH}(t)=27\pi \,G^2 m_{BH}^2(t)\,\rho_{rad}(t).
\end{equation}
\label{6}
This follows from the capture cross section for photons from infinity being equal to $\sigma_{phot}=27\pi\,G^2 m_{BH}^2$ (cf. Ref.~\cite{shapiro83}). For the early times here in question, one has
\begin{equation}
\rho_{rad}=\frac{3H^2}{8\pi G}, \;\; {\rm{with}} \;\; H=\frac{1}{2t}.
\end{equation}
\label{7}
Now put $t=10^{-33}$ s. We obtain $\rho_{rad}=1.9\times 10^{54} \;{\rm{GeV}}^4$, or $4.0\times 10^{92}\; {\rm{erg/cm^3}}$ which, when inserted into Eq.~(6), is conveniently written as
\begin{equation}
\frac{\dot{m}_{BH}}{m_{BH}}=1.4\left( \frac{m_{BH}}{m_{Pl}}\right) \times 10^{23}\; \rm{s^{-1}},
\end{equation}
\label{8}
$m_{Pl}=G^{-1/2}$ being the Planck mass. This is thus the relative rate of mass increase due to accretion. Our condition about mass accretion being negligible, means that the expression (8) must be much smaller than the relative rate of energy loss due to evaporation. The latter is roughly equal to $1/(3\tau)$ according to Eq.~(5). It follows that the condition on $\tau$ becomes
\begin{equation}
\tau \ll \left( \frac{m_{Pl}}{m_{BH}}\right)\times 10^{-24}\; {\rm{s}}.
\end{equation}
\label{9}
This condition will implicitly be assumed below. The greater the value of $m_{BH}$, the stronger becomes the restriction on $\tau$.

Let us also mention another point, related to our assumption that all black holes form almost simultaneously just prior to the instant $t=t_0$. As emphasized by Barrow {\it et al.} \cite{barrow91}, in practice a hole of mass $m_{BH}$ cannot form until the cosmological horizon mass $M_{HOR}(t)$ exceeds $m_{BH}$. Assuming radiation dominance, the horizon mass at time $t$ is given by
\begin{equation}
M_{HOR}(t)=\frac{t}{G}.
\end{equation}
\label{10}
Thus, if we put $t=10^{-33}$ s, we see that the formed black holes must obey the restriction
\begin{equation}
m_{BH} < 4\times 10^5\; {\rm g}.
\end{equation}
\label{11}

\section{The two-fluid system of PBHs and radiation}

In our picture there are two fluids present after the production time $t_0$: one PBH fluid whose decay in the comoving system of coordinates is given by Eq.~(5), and one radiation fluid whose description will be taken to be the same as given in Ref.~\cite{zimdahl98}. Let us first consider the temperature behaviour of the PBH component. To this end we need a bit of relativistic fluid mechanics. We assume that the four-velocity of the cosmic fluid,  $U^\mu=(U^0, U^i)$, is common for the two sub-fluids. In a local inertial rest inertial frame thus $U^0=-U_ 0=1,~U^i=0$.  The general expression for the fluid energy-momentum tensor is, for each sub-fluid component,
\begin{equation}
T^{\mu \nu}=\rho U^\mu U^\nu+(p+\Pi)h^{\mu \nu},
\end{equation}
\label{12}
where $h^{\mu\nu}=g^{\mu\nu}+U^\mu U^\nu$ is the projection tensor, and $\Pi$ the extra pressure brought about by viscosity, by matter production, or eventually by both these effects together. Taking the four-divergence of Eq.~(12) and multiplying with $U_\mu$ we have
\begin{equation}
\dot{\rho}+(\rho+p+ \Pi) \theta=-U_\mu {T^{\mu\nu}}_{;\nu},
\end{equation}
\label{13}
where the scalar expansion is $\theta \equiv       {U^\mu}_{;\mu}=3\dot{a}/a$, $a$ being the scale factor in the FRW line element
\begin{equation}
ds^2=-dt^2+a^2(t)\left( \frac{dr^2}{1-kr^2}+r^2 d\Omega^2 \right).
\end{equation}
\label{14}
For the PBH sub-fluid, $\Pi_{BH}=0$. Neglecting also its pressure, $p_{BH}=0$, we get for the PBH component
\begin{equation}
\dot{\rho}_{BH}+\rho_{BH}\, \theta=-U_\mu\,{T_{BH}^{\mu\nu}}_{;\nu}.
\end{equation}
\label{15}
From Eq.~(2) we have in the local comoving frame \footnote{ As a digression, we note that according to Eq.~(16) there could in principle be a short time period where $\Gamma$ is {\it positive}, corresponding to a production of PBH particles. This production era must obviously be prior to $t_0$. We do not model this era here.}
\begin{equation}
\dot{n}+n\theta=n\Gamma,
\end{equation}
\label{16}
which for the PBH component implies
\begin{equation}
\dot{n}_{BH}+n_{BH}\,\theta=0,
\end{equation}
\label{17}
in view of our assumption (3). Putting $\rho_{BH}=m_{BH}\,n_{BH}$ we obtain from Eqs.~(15) and (17)
\begin{equation}
n_{BH}\,\dot{m}_{BH}=-U_\mu\,{T_{BH}^{\mu\nu}}_{;\nu}.
\end{equation}
\label{18}
Next, we consider the thermodynamical aspects of the system, starting from the thermodynamic identity ($k_B=1$) which holds for each sub-fluid component,
\begin{equation}
nT\,\dot{\sigma}=\dot{\rho}-\frac{\rho+p}{n}\dot{n},
\end{equation}
\label{19}
where $\sigma$ is the nondimensional entropy per particle. We regard $n$ and $T$ as independent variables, so that $\rho=\rho(n,T)$. We differentiate this with respect to $t$, insert $\dot{\rho}$ from Eq.~(13) and $\dot{n}$ from Eq.~(16), and observe the useful relation
\begin{equation}
\rho+p-n\left(\frac{\partial \rho}{\partial n}\right)_T=T\left(\frac{\partial p}{\partial T}\right)_n,
\end{equation}
\label{20}
which follows from the property of $d\sigma$ being an exact differential \cite{weinberg71}. We get
\begin{equation}
\frac{\dot{T}}{T}=-\left( \frac{\partial p}{\partial \rho}\right)_n \theta-\frac{1}{T(\partial \rho/\partial T)_n}\left[ \Pi\,\theta+\left( \frac{\partial \rho}{\partial n}\right)_T \,n\Gamma+U_\mu {T^{\mu\nu}}_{;\nu} \right].
\end{equation}
\label{21}
This is the same form as given in Ref.~\cite{brevik96}, with the addition of the last term, which in turn reflects the fact that we are dealing with a non-closed system. When applied to the PBH component Eq.~(21) yields, when $\Gamma_{BH}=0$,
\begin{equation}
\rho_{BH}\frac{\dot{T}_{BH}}{T_{BH}}=U_\mu\,{T_{BH}^{\mu\nu}}_{;\nu},
\end{equation}
\label{22}
where we have used that $m_{BH}=1/(8\pi GT_{BH})$. 
From Eqs.~(22) and (18),
\begin{equation}
\frac{\dot{T}_{BH}}{T_{BH}}=-\frac{\dot{m}_{BH}}{m_{BH}},
\end{equation}
\label{23}
This is actually the same as obtained in Ref.~\cite{zimdahl98}. The relationship is independent of whether we put $\Gamma_{BH}=-1/\tau$ or $\Gamma_{BH}=0$. It might perhaps seem surprising at first sight that the temperature in the black-hole fluid is not influenced by an eventual (negative) production rate of black holes. However, the present model is based upon a perfect fluid assumption; therefore one should not expect that an existing black hole "feels" if another black hole is created or annihilated. The temperature $T_{BH}$ is associated with each black hole separately, and cannot be thought of as a conventional common temperature for the whole fluid.  

Consider now the entropy four-vector, $S^\mu=n\sigma U^\mu$. This expression holds for each of the components separately. Taking the four-divergence of it we obtain, when using Eqs.~(13),(19) as well as the production equation $(nU^\mu)_{;\mu}=n\Gamma$,
\begin{equation}
{S^\mu}_{;\mu}=-\frac{\Pi}{T}\, \theta-\frac{\mu}{T}\,n\Gamma-\frac{1}{T}U_\mu {T^{\mu\nu}}_{;\nu},
\end{equation}
\label{24}
with $\mu=(\rho+p)/n-T\sigma $ being the Gibbs potential per particle. As ${S^\mu}_{;\mu}=n(\dot{\sigma}+\sigma \Gamma)$ in the local rest frame, we can alternatively express Eq.~(24) in the form
\begin{equation}
\dot{\sigma}=-\frac{\Pi}{nT}\,\theta-\frac{\rho+p}{nT}\,\Gamma-\frac{1}{nT}U_\mu {T^{\mu\nu}}_{;\nu}.
\end{equation}
\label{25}
Again, this is an appropriate generalization of the corresponding expression given in \cite{brevik96}. When applied to the PBH component Eqs.~(24) and (25) yield, in view of $\Pi_{BH}=0,~\Gamma_{BH}=0$ as well as Eq.~(18),
\begin{equation}
S_{BH;\mu}^\mu=n_{BH}\frac{\dot{m}_{BH}}{T_{BH}},
\end{equation}
\label{26}
or
\begin{equation}
\dot{\sigma}_{BH}=\frac{\dot{m}_{BH}}{T_{BH}}.
\end{equation}
\label{27}
Here Eq.~(27) has the form that we would expect. 

We move on to the radiation sub-fluid. With $\Pi_{rad}=0$, and since generally $\mu_{rad}=0$, we first obtain when using Eq.~(24),
\begin{equation}
S_{rad;\mu}^\mu=-\frac{1}{T_{rad}}\,U_\mu T_{rad;\nu}^{\mu\nu}=\frac{1}{T_{rad}}\,U_\mu T_{BH;\nu}^{\mu\nu},
\end{equation}
\label{28}
where in the second equation we have exploited the zero-divergence condition for the total energy-momentum tensor $T_{rad}^{\mu\nu}+T_{BH}^{\mu\nu}$. Thus
\begin{equation}
S_{rad;\mu}^\mu= -n_{BH}\,\frac{\dot{m}_{BH}}{T_{rad}}
\end{equation}
\label{29}
in view of Eq.~(18). From the specific entropy equation (25) we obtain, when setting $p_{rad}=\frac{1}{3}\rho_{rad}$ and assuming  entropy  conservation, $\dot{\sigma}_{rad}=0$,
\begin{equation}
\Gamma_{rad}=-\frac{3}{4}\frac{\rho_{BH}}{\rho_{rad}}\frac{\dot{m}_{BH}}{m_{BH}}.
\end{equation}
\label{30}
Thus, whereas we could above put $\Gamma_{BH}=0$ in the case of pure black-hole evaporation, this is not the case for $\Gamma_{rad}$. When black holes evaporate ($\dot{m}_{BH}<0$), one gets a positive value of $\Gamma_{rad}$ and radiation particles become produced, as one would expect.

Taking the four-divergence of the total entropy four-vector we get from Eqs.~(26) and (29)
\begin{equation}
{S^\mu}_{;\mu} \equiv S_{BH;\mu}^\mu+S_{rad;\mu}^\mu=\rho_{BH}\,\frac{\dot{m}_{BH}}{m_{BH}}\left(\frac{1}{T_{BH}}-\frac{1}{T_{rad}}\right).
\end{equation}
\label{31}
This is an important result. The second law of thermodynamics requires ${S^\mu}_{;\mu}$ to be non-negative, so that the case of pure BH evaporation we must have $T_{BH}>T_{rad}$. Black hole evaporation is connected with entropy production. If the converse were true in some time period, $T_{BH}<T_{rad}$, then the second law would require that $\dot{m}_{BH}>0$. The black holes would accordingly increase in mass during such a period.

For comparison we write down some expressions obtained on the basis of assuming instead $\Gamma_{BH}=-1/\tau$, as in the ZP paper. Instead of Eqs.~(30) and (31) one gets
\begin{equation}
\Gamma_{rad|ZP}=-\frac{3}{4}\,\frac{\rho_{BH}}{\rho_{rad}}\left( -\frac{1}{\tau}+\frac{\dot{m}_{BH}}{m_{BH}}\right),
\end{equation}
\label{32}
\begin{equation}
{S^\mu}_{;\mu |ZP}=-\frac{\rho_{BH}}{\tau}\left(\frac{\sigma_{BH}}{m_{BH}}-\frac{1}{T_{rad}}\right)+
\rho_{BH}\,\frac{\dot{m}_{BH}}{m_{BH}}\left(\frac{1}{T_{BH}}-\frac{1}{T_{rad}}\right).
\end{equation}
\label{33}
The structure is here seen to be more complicated, especially for the entropy four-divergence. 

We will need also the expression for the relative rate of change of the temperature for photons. From the general expression 
\begin{equation}
\frac{\dot{T}_{rad}}{T_{rad}}=-\frac{1}{3}(\theta-\Gamma_{rad})
\end{equation}
\label{34}
we get, assuming $\Gamma_{BH}=0$,
\begin{equation}
\frac{\dot{T}_{rad}}{T_{rad}}=-\frac{1}{3}\left( \theta+\frac{3}{4}\,\frac{\rho_{BH}}{\rho_{rad}}\frac{\dot{m}_{BH}}{m_{BH}}\right),
\end{equation}
\label{35}
whereas assuming $\Gamma_{BH}=-1/\tau$ we get
\begin{equation}
\frac{\dot{T}_{rad|ZP}}{T_{rad|ZP}}=-\frac{1}{3}\left[ \theta+\frac{3}{4}\frac{\rho_{BH}}{\rho_{rad}}\left(-\frac{1}{\tau}+
\frac{\dot{m}_{BH}}{m_{BH}}\right) \right].
\end{equation}
\label{36}

\subsection{What happens for $t \gtrsim t_0$?}

We are now able to give a physical interpretation of the two-component cosmic fluid near the production time $t=t_0$. As mentioned above, we associate $t_0$ with the end of the inflationary era. At this instant we will moreover make the explicit assumption that the {\it sub-fluids have the same temperature}. That is, the PBHs are taken to be created with the same temperature as the radiation field which they were created from. Since the two fluid temperatures coincide, Eq.~(31) tells that no entropy will be produced at $t_0$. Recall that at $t_0$, the second law of thermodynamics permits black-hole mass accretion as well as evaporation to occur. The first case corresponds to $\dot{m}_{BH}>0,\, T_{BH}<T_{rad}$, the second case corresponds to $\dot{m}_{BH}<0,\, T_{BH}>T_{rad}$. 

At this point we have to make a digression, to check that our assumption about equality of the two sub-fluid temperatures at $t=t_0$ makes physical sense. Evidently, the produced black holes have to be smaller than the universe itself. At time $t$ the particle horizon $R_p(t)$, i.e., the radius of the observable part of the universe, is
\begin{equation}
R_p(t)=a(t)\int_0^t \frac{dt'}{a(t')}=2t=H^{-1},
\end{equation}
\label{37}
where we have used that $a(t) \propto t^{1/2}$. At $t=10^{-33}$ s thus $R_p=6\times 10^{-23}$ cm. As mentioned above, $\rho_{rad}=4.0\times 10^{92}\; {\rm{erg/cm^3}}$, and the temperature $T_{rad}$ can be found from the equation $\rho_{rad}=a_{rad}T_{rad}^4$, where $a_{rad}=7.56\times 10^{-15}\; {\rm{erg/(cm^3\,K^4)}}$ is the radiation constant. We find $T_{rad}=4.8\times 10^{26}$ K, or $4.1\times 10^{13}$ GeV. The Schwarzschild radius of a black hole is $r_{BH}=2G m_{BH}=1/(4\pi T_{BH})$  and we get, when assuming $T_{rad}=T_{BH}$, $r_{BH}=3.8\times 10^{-29}$ cm. This is seen to be much smaller than $R_p$. There is thus ample room for the BH gas in the universe, showing that our assumption about equal-temperature of the sub-fluids at $t=t_0$ is consistent.

Assume next that $\dot{m}_{BH}=0$ at $t_0$. Then no process at all, except from the expansion of the universe, will run. According to Eq.~(23), $T_{BH}$ will then stay constant, whereas $T_{rad}$ will drop due to the expansion. That means, for $t>t_0$ one has $T_{rad}<T_{BH}$, and an evaporation process has to start, for the second law of thermodynamics not to be violated.

If on the other hand an evaporation process is already going on at $t_0$, the black hole temperature $T_{BH}$ will have to rise according to Eq.~(23). Equation (35) then tells that the radiation temperature $T_{rad}$ may rise or fall depending on the ratio $\rho_{BH}/\rho_{rad}$. If the process of pure evaporation is to be maintained, the radiation temperature must not rise faster than the black-hole temperature. From Eqs.~(23), (35), and (5), this leads to the following condition at $t_0$:
\begin{equation}
\beta \leq 4(\theta_0 \tau+1),
\end{equation}
\label{38}
where we have defined $\beta$ by
\begin{equation}
\beta=\frac{\rho_{BH}(t_0)}{\rho_{rad}(t_0)},
\end{equation}
\label{39}
$\theta_0$ meaning the scalar expansion at $t_0$. If the inequality (38) is broken, the radiation temperature will at some time exceed that of the black holes, and the evaporation process will no longer be thermodynamically allowed. The maximum value of $\beta$ represents the largest density ratio for which evaporation can take place. The possible values of $\beta$ depend on the expansion and the life time of the black holes. The larger the value of $\tau$, the larger may the initial abundance of the black holes be. This is physically reasonable: a long life time means that the black hole is large and evaporates less intensively than a small black hole. We should therefore expect that large black holes contribute less to the heating of the radiation sub-fluid than small black holes do. Recall, however, the restriction on the magnitude of $\tau$ that lies in Eq.~(9).

It is instructive to get an idea about the typical magnitude of the quantity $\theta_0 \tau$. Let us assume, as an example, that black holes of mass $m_{BH}=100 $ g $ \approx 0.5\times 10^7 \; m_{Pl}$ are formed at $t=10^{-33}$ s. (This mass satisfies the condition (11).) From Eq.~(9) we then have $\tau \ll 2\times 10^{-31}$ s, in order to permit neglect of accretion.  The scalar expansion is $\theta_0 =3H=3/(2t)=1.5\times 10^{33}\; {\rm s^{-1}}$, so that we obtain $\theta_0\tau \ll 300$.

\subsection{Reheating temperature}

We assume henceforth the process of pure evaporation, so that for $t>t_0$ the black-hole temperature increases and stays above the radiation temperature. The decay rate, Eq.~(5), will then hold for the entire period $t>t_0$.

As the second law of thermodynamics limits the reheating, a reheating temperature $T_{rad}^{reh}$ may be defined. It is the maximum temperature obtained for the radiation sub-fluid, and corresponds to $\dot{T}_{rad}=0$. Using the rate formula (35), the equation of state $\rho_{rad}=3n_{rad}\,T_{rad}$, and Eq.~(5), we find , when assuming the reheating to take place immediately after $t=t_0$:
\begin{equation}
T_{rad}^{reh}=\frac{1}{12}\,\frac{n_{BH}(t_0)}{n_{rad}(t_0)}\,\frac{m_{BH}(t_0)}{\theta_0 \tau}.
\end{equation}
\label{40}
By contrast, ZP obtained a value four times greater:
\begin{equation}
T_{rad|ZP}^{reh}=4\,T_{rad}^{reh}.
\end{equation}
\label{41}
This difference is physically understandable from a comparison between Eqs.~(32) and (30). In the ZP case there is one extra term proportional to $1/\tau$, contributing to the reheating. One would therefore expect that a higher amount of energy exists in the radiation field before the second law of thermodynamics forces the heating of the radiation to terminate.

\subsection{Equilibrium solutions}

Some years ago Barrow {\it et al.} \cite{barrow91} found cosmological solutions for which the ratio of the energy density of the black hole component to the energy density of the radiation component remained equal to a constant $\kappa$,
\begin{equation}
\rho_{BH}=\kappa \rho_{rad}.
\end{equation}
\label{42}
Solutions of this kind are called equilibrium solutions.

The question naturally emerges: are such solutions possible in our model? To investigate this point, let us first observe that Eqs.~(15) and (18) provide us with the following equation in the case of pure evaporation:
\begin{equation}
\dot{\rho}_{BH}+\rho_{BH}\,\theta=n_{BH}\,\dot{m}_{BH}.
\end{equation}
\label{43}
The corresponding equation for the radiation component reads
\begin{equation}
\dot{\rho}_{rad}+\frac{4}{3}\rho_{rad}\,\theta=-n_{BH}\,\dot{m}_{BH}.
\end{equation}
\label{44}
Inserting Eq.~(42) into Eq.~(43) and making use of Eq.~(30) we obtain
\begin{equation}
\Gamma_{rad}=-\frac{3}{4}\kappa \left(\theta+\frac{\dot{\rho}_{rad}}{\rho_{rad}} \right).
\end{equation}
\label{45}
With $\rho_{rad}=3n_{rad}\,T_{rad}$ the last term to the right becomes
\begin{equation}
\frac{\dot{\rho}_{rad}}{\rho_{rad}}=\frac{\dot{n}_{rad}}{n_{rad}}+\frac{\dot{T}_{rad}}{T_{rad}}=\frac{4}{3}(\Gamma_{rad}-\theta),
\end{equation}
\label{46}
in view of Eqs.~(16) and (34). Altogether we thus obtain, when $\Gamma_{BH}=0$,
\begin{equation}
\Gamma_{rad}=\frac{1}{4}\frac{\kappa}{\kappa+1}\,\theta.
\end{equation}
\label{47}
This shows that for an equilibrium solution to exist, $\Gamma_{rad}$ has to be proportional to $\theta$. Since $\Gamma_{rad}=-(3\kappa /4)\dot{m}_{BH}/m_{BH}$, this implies in turn that $\dot{m}_{BH}/m_{BH}$ has to be proportional to $\theta$, or
\begin{equation}
 \theta \propto \left( 1-\frac{t-t_0}{\tau} \right)^{-1}, 
\end{equation}
\label{48}
according to Eq.~(5). A relationship of this form is physically unlikely. Such a time dependence for $\theta$ is quite different from that given, for instance, in \cite{brevik94} (for a viscous universe). We conclude that equilibrium solutions do not lead to a realistic space-time dynamics when $\Gamma_{BH}=0$ is assumed. The reason for this deviation from the theory of Barrow {\it et al.} is easy to trace out: these authors consider a very different kind of PBH model, in which the initial number of holes between $m$ and $m+dm$
 is taken to be proportional to $m^{-n}\,dm$,
 with $n$ (the spectral index) lying between 2 and 3. This evidently contrasts our adopted model where all the PBHs have initially the same mass $m_{BH}(t_0)$. We have no reason to expect that the consequences of two so different models are in agreement.

The possibility of equilibrium solutions was investigated also by ZP: they found that by assuming $\Gamma_{BH}=-1/\tau$, equilibrium solutions were  mathematically impossible. The qualitative agreement between the consquences of 
the ZP theory and the closely related present theory is of course what we would expect.

\subsection{Time dependence of the densities}

Integrating Eq.~(43) from $t_0$ to $t$, inserting $\theta=3\dot{a}(t)/a(t)$ and using Eq.~(5), we get
\begin{equation}
\rho_{BH}(t)=\rho_{BH}(t_0)\,\left(\frac{a_0}{a}\right)^3 \left(1-\frac{t-t_0}{\tau}\right)^{1/3}.
\end{equation}
\label{49}
Equation (44) for the radiation now becomes
\begin{equation}
\dot{\rho}_{rad}+4\frac{\dot{a}}{a} \,\rho_{rad}=
\frac{\rho_{BH}(t_0)}{3\tau}\left(\frac{a_0}{a}\right)^3 \left(1-\frac{t-t_0}{\tau}\right)^{-2/3}.
\end{equation}
\label{50}
We multiply this equation with $a^4$ and integrate:
\begin{equation}
\rho_{rad}(t)=\rho_{rad}(t_0)\frac{a_0^4}{a^4}+
\frac{\rho_{BH}(t_0)}{3\tau}\frac{a_0^3}{a^4}\int_{t_0}^t a(t')\left(1-\frac{t'-t_0}{\tau}\right)^{-2/3}dt'.
\end{equation}
\label{51}
Knowledge about $\rho_{rad}(t_0)$ and $\rho_{BH}(t_0)$ plus solution of the field equations for $a(t)$ thus determine the dynamics of the entire system. In particular, for a flat FRW universe the first Friedmann equation yields
\begin{equation}
\frac{\dot{a}^2}{a^2}=\frac{8\pi G}{3}(\rho_{BH}+\rho_{rad}).
\end{equation}
\label{52} 
Equations (49) and (51) are simpler than the corresponding equations obtained by ZP by assuming nonvanishing $\Gamma_{BH}$.

\section{On an effective one-component model}

Instead of working in terms of a two-component fluid model let us now try to formulate an effective one-component model. As mentioned earlier, this is analogous to treating water with air bubbles as one single fluid. The transformation from a two-fluid description to an effective one-fluid description, in the form worked out by Zimdahl \cite{zimdahl97}, is not directly applicable here since there exists a temperature difference between the two components: the evaporating black holes become hotter, while the radiation may cool down due to the expansion of the universe. We may nevertheless introduce an "equilibrium" temperature $T$, but evidently so only in a formal sense except in the initial stages near $t=t_0$.

As we will see, the lack of a physical equilibrium temperature does not prevent us from constructing in some sense an effective one-component theory. A key element in the formalism is that the evaporation process is represented in terms of a viscous pressure. As above, we assume henceforth that $\Gamma_{BH}=0$.

Following the ZP paper we start from the Gibbs equation for a two-fluid model close to equilibrium:
\begin{equation}
T\,d\sigma= d\left(\frac{\rho}{n}\right)+pd\left(\frac{1}{n}\right)-(\mu_{BH}-\mu_{rad})d\left(\frac{n_{BH}}{n}\right),
\end{equation}
\label{53}
where $T$ is the equilibrium temperature for the system of a conventional fluid and a PBH component. In our case the equations of state are $p=p_{rad}(n_{rad},T)$ and $\rho=\rho(n, n_{BH}, T)$, with $n=n_{BH}+n_{rad}$. We {\it define} the "equilibrium" temperature $T$ through the equation
\begin{equation}
\rho_{BH}(n_{BH}, T_{BH})+\rho_{rad}(n_{rad},T_{rad})=\rho(n, n_{BH},T).
\end{equation}
\label{54}
Since moreover $\mu_{rad}=0$ we have
\begin{equation}
n\mu=n_{BH}\,\mu_{BH},\quad n\Gamma=n_{rad}\,\Gamma_{rad},
\end{equation}
\label{55}
\begin{equation}
n\sigma=n_{BH}\,\sigma_{BH}+n_{rad}\,\sigma_{rad},
\end{equation}
\label{56}
whereas upon use of $p_{BH}=0,\, \sigma_{BH}=4\pi G m_{BH}^2,\,T_{BH}=1/(8\pi G m_{BH})$ in the relation $\mu_{BH}=m_{BH}-\sigma_{BH}\,T_{BH}$ we get
\begin{equation}
\mu_{BH}=\frac{1}{2}m_{BH},\quad n\mu=\frac{1}{2}\rho_{BH}.
\end{equation}
\label{57}
The energy-momentum tensor is given by Eq.~(12) as before, whereas Eq.~(13) because of conservation of total energy is replaced by
\begin{equation}
\dot{\rho}+(\rho+p+\Pi)\theta=0
\end{equation}
\label{58}
in the effective one-component description. From Eq.~(53) we now obtain, after a brief calculation observing that $(n/n_{rad})\Gamma=\Gamma_{rad}$,
\begin{equation}
n\dot{\sigma}=-\frac{\theta}{T}\,\Pi+\left(n_{BH}\,\frac{\mu_{BH}}{T}-\frac{\rho+p}{T}\right)\Gamma.
\end{equation}
\label{59}
By means of Eqs.~(57) we get the following expression for the four-divergence ${S^\mu}_{;\mu}=n(\dot{\sigma}+\sigma \Gamma)$ of the effective entropy four-vector:
\begin{equation}
{S^\mu}_{;\mu}=-\frac{\theta}{T}\,\Pi.
\end{equation}
\label{60}
This formula has a remarkably simple appearance, since the effective production rate $\Gamma$ is absent. The whole of the entropy production is attributable to the excess pressure $\Pi$, which for a conventional viscous fluid is expressed as $\Pi=-\zeta \theta$, $\zeta$ being the bulk viscosity. Thus as far as entropy production is concerned, in our model the universe, composed of radiation and non-relativistic evaporating PBHs at a higher temperature than the radiation, can be described in terms of an effective bulk viscosity.

The form of Eq.~(60) relies upon the condition $\Gamma_{BH}=0$. The corresponding Eq.~(68) in the ZP paper is more complicated.

We may also compare the two expressions obtained for the production of entropy. We recall that in the two-component model, expression (31) was derived on basis of the assumption about black hole evaporation. Comparing this expression with expression (60), we get
\begin{equation}
\Pi=\frac{4}{3}\frac{T}{\theta}\,\rho_{rad}\,\Gamma_{rad}\left(\frac{1}{T_{BH}}-\frac{1}{T_{rad}}\right),
\end{equation}
\label{61}
where Eq.~(30) has been taken into account. As we noted earlier, pure evaporation corresponds to $T_{BH}>T_{rad}$, so that $\Pi$ has to be negative according to Eq.~(61) since $\Gamma_{rad} >0$. (In other words, $\zeta =-\Pi/\theta$ has to be positive, as it should.)

Summing up, we see that the legitimacy of the one-component model rests upon the possibility that we have to put the temperatures of the two sub-fluids equal, at $t=t_0$. Although some of results of this theory are simple, in particular the expression (60) for the entropy, we think on the whole that the one-fluid model is from a physical point of view inferior to the two-fluid model considered earlier. The "equilibrium" temperature $T$ defined in Eq.~(54) is after all, similarly as in the ZP paper, introduced in a formal sense, so that the main physical relevance of the one-fluid model appears to be in the period just after $t=t_0$, when the two temperatures have not got the  opportunity to separate significantly.

\section{Concluding remarks}

Our main new contribution in the present paper has been to provide an analysis of cosmological BH theory when the production rate $\Gamma_{BH}$ is assumed to be zero throughout. This assumption is based upon both physical and mathematical considerations. Physically, the interpretation of a production rate in the form $ \Gamma_{BH}=-1/\tau$, as assumed in the ZP paper \cite{zimdahl98}, is problematic (we made an attempt to formulate a natural interpretation in Sec. 1), whereas mathematically the condition $\Gamma_{BH}=0$ leads to considerable simplifications. In particular, the expression (31) for the four-divergence of the entropy four-vector fits in with the second law of thermodynamics in a simple way,  showing how the evaporation of the PBHs leads to a positive entropy production. In principle, a process of this sort may even be one of the reasons for the remarkably high specific entropy, $\sigma \simeq 10^9$, in the universe. 

Another  basic assumption in our theory is that all the PBHs have the same mass. Physically, this is consistent with assuming that all the PBHs were formed during a short time interval just preceding the instant $t=t_0$, in the early universe. The time $t_0$ is naturally taken to be at the end of the inflationary era. Immediately before $t_0$, the PBHs are assumed to be formed at the radiation temperature. Immediately after $t_0$, the PBHs may evaporate freely and the thermal equilibrium between black holes and radiation is lost.

When the PBHs are evaporating for $t \geq t_0$, the radiation temperature $T_{rad}$ cannot rise faster than then black.hole temperature The reheating temperature is given by Eq.~(40), and is four times higher than the reheating temperature (41) obtained in the ZP formulation.

It should be noted that our theory neglects the accretion effect for the black holes. The mathematical condition for this is given by Eq.~(9).

When considering instead an effective one-component model, the simple appearance of the four-divergence formula (60), again being a consequence of $\Gamma_{BH}=0$, should be noticed. This formula shows how the entropy production can be ascribed to the existence of an effective  bulk viscosity $\zeta=-\Pi/\theta$ in the mixed, common  fluid. However, on the whole, the one-component model seems to be less motivated physically than a two-component model. The difference in temperature between the BH component and the radiation component signalizes a physical instability which is best handled physically when one regards the cosmic fluid as two separate sub-fluids.

Finally, to widen the scope of applicability of the above formalism,we mention another interpretation of it that also appears to be physically possible. Instead of requiring $\Gamma_{BH}$ to be strictly zero, as we have done above, one can assume that this requirement is only as {\it approximation} to the real physics. In that case, the equation $\Gamma_{BH}=0$ would correspond to a neglect of possible remnants of the PBHs. This would restrict the above formalism to hold only in the restricted time range $t-t_0 <\tau$.

\end{document}